\documentclass[12pt]{article}
\usepackage{amssymb,amsmath,epsfig}
\usepackage{graphicx}
\allowdisplaybreaks
\begin{document}
\title{\bf Compact Objects admitting Finch-Skea Symmetry in $f(\mathcal{R},T^2)$ Gravity}
\author{M. Sharif \thanks {msharif.math@pu.edu.pk} and Sana Manzoor
\thanks{sanamanzoormath@gmail.com}\\
Department of Mathematics and Statistics,\\
The University of Lahore 54000, Lahore-Pakistan.}
\date{}
\maketitle

\begin{abstract}
This paper investigates the viability and stability of anisotropic
compact stars in the framework of $f(\mathcal{R},\mathrm{T}^{2})$
theory ($\mathcal{R}$ is the Ricci scalar and
$\mathrm{T}^{2}=\mathrm{T}_{\tau\upsilon}\mathrm{T}^{\tau\upsilon}$).
In this perspective, we use Finch-Skea symmetry and consider
different $f(\mathcal{R},\mathrm{T}^{2})$ models to examine physical
characteristics of compact stars. We match interior spacetime with
the exterior region to find the values of unknown constants.
Further, we analyze the behavior of several physical quantities such
as effective matter variables, energy bounds, anisotropic factor and
equation of state parameters in the interior of Vela X-1, SAX J
1808.4-3658, Her X-1 and PSR J0348+0432 stars. The equilibrium state
of these compact stars is examined through the
Tolman-Oppenheimer-Volkoff equation and their stability is checked
by causality condition, Herrera cracking approach and adiabatic
index. It is found that all the required conditions are fulfilled
for the considered models. We can conclude that more viable and
stable compact stars are obtained in this modified theory.
\end{abstract}

\section{Introduction}

The astrophysical and cosmological observations inspired many
cosmologists in recent years to discuss the cosmos and its enigmatic
components. Stars are assumed as the fundamental elements of
astronomy and the basic building blocks of galaxies. The evolution
of stars and planets are significantly influenced by the fusion
processes. A star maintains its equilibrium position by balancing
the inward force of gravity with the outward pressure produced by
the fusion process. Once the nuclear fuel is consumed, there does
not seem enough pressure to prevent the star from collapsing. This
process yields heat and light in space as well as the formation of
new compact objects such as \emph{white dwarfs}, \emph{neutron
stars} and \emph{black holes} based on their original mass before
collapsing.

Compact stars are considered as the main constituents of massive
stars and have attracted many researchers. In this regard, neutron
stars have attained much attention because of their unique
characteristics and configurations. Baade and Zwicky \cite{1}
examined the geometry of compact objects and gave the idea of
neutron stars. The rotating neutron stars (pulsars) such as Her X-1
indicates the existence of neutron stars. The concept of neutron
stars received observational confirmation after the discovery of
pulsars \cite{2}. The physical characteristics of pulsars with
various matter configurations have widely been investigated in the
literature. Dev and Gleiser \cite{3} examined the physical behavior
of pulsars from several perspectives. Mak and Harko \cite{4} used
mass-radius relation to analyze the stability of pulsars. Kalam et
al \cite{5} explored the viable and stable compact stars by using
Karori-Barua solutions. Hossein et al \cite{6} used Krori-Barua
solutions with cosmological constant to investigate the stability of
anisotropic objects. Maurya et al \cite{7} found anisotropic analog
of the Durgapal and Fuloria perfect fluid solution and examined the
validity of the model. Singh and Pant \cite{8} obtained a family of
new exact solutions for relativistic anisotropic stellar objects by
considering Karmarkar conditions.

The anisotropy in spherically symmetric objects affects some
important physical features of relativistic objects. Ruderman
\cite{11} suggested that if the matter density of relativistic
objects is equal to $10^{15}g/cm^{3}$ then nuclear matter exhibits
anisotropy. The distribution of matter exhibits pressure anisotropy
due to the presence of phase transition, viscosity, pion
condensation and super fluid \cite{12}. Bowers and Liang \cite{13}
presented physical properties of anisotropic pressure and discussed
the anisotropy of a relativistic sphere. In various self-gravitating
systems, Herrera and Santos \cite{14} studied consequences of local
anisotropy. Hernandez and Nunez \cite{15} examined the anisotropic
solution and equilibrium composition of compact stellar objects.
Kalam et al \cite{17} used Karori-Barua solutions to examine the
behavior of anisotropic quark stars. Paul and Deb \cite{18}
discussed the effect of anisotropy on the geometry of pulsars.

Dourah and Ray \cite{19} studied the metric solutions for compact
stars. Later, it was noticed that these solutions were not suitable
for compact stars. Finch and Skea \cite{20} formulated a new metric
for spherical symmetric compact stellar configuration. Afterwards,
the Finch-Skea metric was modified in four dimensions to derive
anisotropic star models \cite{17,21,23}. Sharma and Ratanpal
\cite{24} proposed a relativistic star model using the Finch-Skea
metric solutions. Bhar \cite{25} calculated physical properties of
compact stars using Finch-Skea metric solutions and Chaplygin
equation of state (EoS) parameter. Pandya et al \cite{26} discussed
anisotropic stellar configuration through Finch-Skea metric
potentials. Bhar et al \cite{9} established a new well-ordered
charged anisotropic solution through this symmetry. Shamir et al
\cite{48} examined the anisotropic compact objects configuration by
applying Finch-Skea solutions.

In the recent years, different cosmic studies have been presented to
understand the beginning of the universe. One of the most accepted
proposals is the big-bang theory which is a remarkable framework to
describe the cosmic evolutionary processes. According to this
theory, all the matter in the cosmos expanded from a single point,
referred to as a singularity. This theory explains the origin of the
universe but it suffers many cosmic issues such as flatness and
horizon problems. In order to resolve these issues, Katrici and
Kavuk \cite{34} generalized $f(\mathcal{R})$ theory by introducing
self-contraction of the energy-momentum tensor
$(T^{2}=T_{\mu\nu}T^{\mu\nu})$ in the functional action, named as
$f(\mathcal{R},T^{2})$ theory which is also called energy-momentum
squared gravity (EMSG). This modified theory resolves the big-bang
singularity and provides a better explanation of current cosmic
accelerated expansion. The cosmological constant resolves the
big-bang singularity by providing the repulsive force in the
background of this theory. This theory follows the true sequence of
cosmological eras and effectively describes cosmic behavior. They
found the Newtonian limit of the model to calculate the extra
acceleration which can affect the perihelion of Mercury. There is a
deviation from the general relativistic (GR) result unless the
energy density of the fluid is constant. The field equations involve
squared and product components of matter variables, which are useful
in studying different cosmological scenarios.

Nari and Roshan \cite{34a} found that EMSG depends on the central
pressure of the star and magnitude of the free parameter $\eta$
which can lead to larger or smaller masses for neutron stars as
compared to GR. This fact is satisfactory in the sense that there
are difficulties in GR for explaining the internal structure of
massive neutron stars, especially their high mass, using ordinary
equations of state \cite{34b}. Since EMSG is introduced to resolve
the singularities, it is natural to expect its deviations from GR
that appear only at high curvature regimes. Therefore, it is also
necessary to investigate it inside compact stars where the energy
scale is high enough to see EMSG deviations from GR. Energy-momentum
squared gravity can help to explain the process of gravitational
collapse in compact objects such as black holes and neutron stars.
This modified theory can capture these nonlinear effects, providing
a more accurate description of the gravitational collapse process.
Compact objects such as black holes are often associated with
singularities. Energy-momentum squared gravity can help to resolve
these singularities by incorporating higher-order corrections to the
energy-momentum tensor. This can lead to a more complete and
physically realistic description of black holes and other compact
objects.

Board and Barrow \cite{35} examined the features of exact solutions
via several cosmic parameters in the perspective of EMSG. Moraes and
Sahoo \cite{37} studies the viable wormhole solutions while Akarsu
et al \cite{38} estimated the mass-radius relation of neutron stars
and explored viable limitations from neutron stars in this gravity.
Bahamonde \cite{39} investigated different models and concluded that
this theory represents current evolution and acceleration of the
universe. Sharif and his collaborators studied the stability of
Einstein universe \cite{40}, viable cosmological solutions through
Noether symmetry approach \cite{40a} and gravastar structures
\cite{40b} in the context of EMSG. Rudra and Pourhassan \cite{10}
examined thermodynamic properties of the universe in the background
of the generalized EMSG.

Many researchers have studied physical attributes of compact objects
in the background of modified gravitational theories. Astashenok et
al \cite{10a} examined the effect of $f(\mathcal{R})$ gravity model
on the stable structure of neutron stars by evaluating their density
and pressure profiles. Shamir and Zia \cite{10b} examined the impact
of electromagnetic field on the anisotropic Her X-1 pulsar in
$f(\mathcal{R},G)$ gravity. Rahaman et al \cite{10c} studied the
existence of anisotropic compact spherical systems admitting
Karmarkar constraint in $f(\mathcal{R},T)$ theory. Sharif and Ramzan
\cite{10d} discussed the nature of different physical parameters and
stability of different stellar objects by considering this
constraint in $f(G)$ theory. Some astrophysical objects such as
neutron stars were examined in the background of EMSG \cite{10e}.
Different symmetries are used in modified theories of gravity to
check the consistency with observational data, addressing
theoretical problems and simplifying the mathematical analysis of
the theory. In this context, we have used Finch-Skea symmetry in the
background of modified $f(\mathcal{R},T^{2})$ theory. Finch-Skea
symmetry is physically viable and non-singular configuration.
Finch-Skea geometry was modified to discuss isotropic as well as
anisotropic stars \cite{10f}. A number of research work in
relativistic astrophysics using Finch-Skea metric has been done in
various modified theories \cite{10g}-\cite{10m}. Sharif and Gul
\cite{10n} investigated the geometry of compact stars admitting
Noether symmetry technique in $f(\mathcal{R},\mathrm{T}^{2})$
theory. They also analyzed the dynamics of gravitational collapse
with different matter distributions and concluded that the modified
terms reduce the collapse rate \cite{10p}.

In this paper, we analyze the viable and stable compact stars
admitting Finch-Skea symmetry in $f(\mathcal{R},\mathrm{T}^{2})$
theory. The paper is planned as follows. In section \textbf{2}, we
explore the equations of motion associated with Finch-Skea symmetry
and determine the values of constants by smooth matching of the
interior (static spherically symmetric) and exterior (Schwarzschild)
metrics. Section \textbf{3} provides viable models of this theory.
In section \textbf{4}, we examine the influence of some significant
physical parameters in the interior of compact stars. Section
\textbf{5} investigates the equilibrium state of the proposed stars
via Tolman-Oppenheimer-Volkoff (TOV) equation and their stability
are examined by the speed of sound and adiabatic index. The final
section provides summary of the results.

\section{$f(\mathcal{R},\mathrm{T}^{2})$ Gravity}

The action of EMSG is defined as \cite{49}
\begin{equation}\label{1}
I=\frac{1}{2\kappa}\int \sqrt{-g}f(\mathcal{R},\mathrm{T}^{2})
d^{4}x+\int \sqrt{-g}L_{m}d^{4}x,
\end{equation}
where determinant of the line element and coupling constant are
denoted by $g$ and $\kappa$, respectively. The correspondent field
equations are
\begin{equation}\label{2}
\mathcal{R}_{\tau\upsilon}f_{\mathcal{R}}+g
_{\tau\upsilon}\nabla_{\tau}
\nabla^{\tau}f_{\mathcal{R}}-\nabla_{\tau}
\nabla_{\upsilon}f_{\mathcal{R}}-\frac{1}{2}g_{\tau\upsilon}f=T_{\tau\upsilon}-\Theta_{\tau
\upsilon}f_{T^{2}},
\end{equation}
where $f=f(\mathcal{R},\mathrm{T}^{2})$, $f_{T^{2}}=\frac{\partial
f}{\partial T^{2}}$, $f_{\mathcal{R}}=\frac{\partial
f}{\partial\mathcal{R}}$ and
\begin{equation}\label{3}
\Theta_{\tau\upsilon}=-2L_{m}(T_{\tau
\upsilon}-\frac{1}{2}g_{\tau\upsilon}T)-\frac{4
\partial^{2}L_{m}}{\partial g^{\tau\upsilon}\partial
g^{\mu\nu}}T^{\mu\nu}-TT_{\tau\upsilon}+2T^{\mu}_{\tau}T_{\upsilon\mu}.
\end{equation}
Rearranging Eq.(\ref{2}), we obtain
\begin{equation}\label{4}
G_{\tau\upsilon}=\mathcal{R}_{\tau\upsilon}-\frac{1}{2}\mathcal{R}g_{\tau
\upsilon}=T^{eff}_{\tau\upsilon},
\end{equation}
where $T^{eff}_{\tau\upsilon}$ are the effective terms of EMSG,
given as
\begin{equation}\label{5}
T^{eff}_{\tau\upsilon}=\frac{1}{f_{\mathcal{R}}}\big[T_{\tau
\upsilon}+\nabla_{\tau}\nabla_{\upsilon}f_{\mathcal{R}}-g_{\tau
\upsilon}\nabla_{\tau}\nabla^{\tau}f_{\mathcal{R}}+\frac{1}{2}g_{\tau
\upsilon}(f-\mathcal{R}f_{\mathcal{R}})-\Theta_{\tau
\upsilon}f_{T^{2}}\big].
\end{equation}
We take a static spherical line element to study the geometry of
compact stars
\begin{equation}\label{6}
ds^{2}_{-}=-e^{\alpha(r)}dt^{2}+e^{\beta(r)}dr^{2}+r^{2}d\Omega^{2},
\end{equation}
where $d\Omega^{2}=d\theta^{2}+\sin^{2}\theta d\phi^{2}$. We assume
anisotropic matter configuration as
\begin{equation}\label{7}
T_{\tau\upsilon}=\rho U_{\tau}U_{\upsilon}+P_{r}
V_{\tau}V_{\upsilon}+P_{t}(U_{\tau}U_{\upsilon}-V_{\tau}V_{\upsilon}+g_{\tau\upsilon}),
\end{equation}
where $U_{\tau}$, $V_{\tau}$, $\rho$, $P_{r}$ and $P_{t}$ are used
as symbols of four-velocity, four-vector, energy density, radial and
tangential pressures, respectively. Consequently, the field
equations turn out to be
\begin{eqnarray}\nonumber
\rho^{eff}&=&\frac{1}{f_{\mathcal{R}}}\bigg[\rho-\frac{f-\mathcal{R}
f_{\mathcal{R}}}{2}+f_{T^{2}}
\big\{L_{m}(\rho+2P_{t}+P_{r})+\rho^{2}+2\rho P_{t}+\rho P_{r}\big\}
\\\label{8}
&+&\frac{1}{e^{\beta}}\bigg\{f''_{\mathcal{R}}-\bigg(\frac{\beta'}{2}
-\frac{2}{r}\bigg)f'_{\mathcal{R}}\bigg\}\bigg],
\\\nonumber
P^{eff}_{r}&=&\frac{1}{f_{\mathcal{R}}}\bigg[p_{r}+\frac{f-\mathcal{R}
f_{\mathcal{R}}}{2}+\big\{L_{m}(\rho-2P_{t}+P_{r})-(P^{2}_{r}-2P_{r}P_{t}+\rho
P_{r})\big\}f_{T^{2}}
\\\label{9}
&-&\frac{1}{e^{\beta}}\bigg(\frac{2}{r}+\frac{\alpha'}{2}\bigg)
f'_{\mathcal{R}}\bigg],
\\\nonumber
P^{eff}_{t}&=&\frac{1}{f_{\mathcal{R}}}\bigg[P_{t}+\frac{f-\mathcal{R}
f_{\mathcal{R}}}{2}+\big\{L_{m}(-P_{r}+\rho)+P_{r}P_{t}-\rho
P_{t}\big\}f_{T^{2}}
\\\label{10}
&-&\frac{1}{e^{\beta}}\bigg\{f'_{\mathcal{R}}\bigg(\frac{\alpha'}{2}
+\frac{1}{r}-\frac{\beta'}{2}\bigg)+f''_{\mathcal{R}}
\bigg\}\bigg].
\end{eqnarray}
\begin{figure}
\epsfig{file=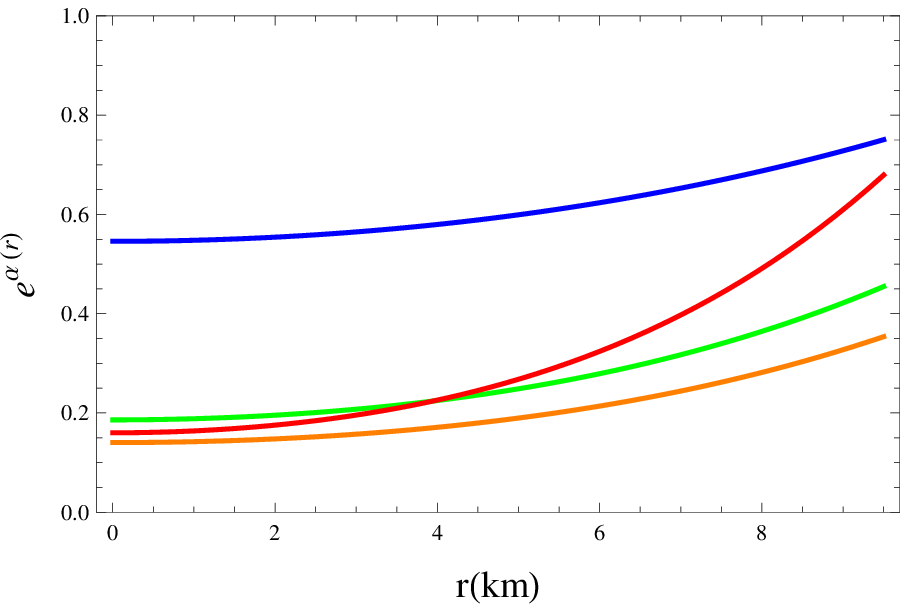,width=.5\linewidth}
\epsfig{file=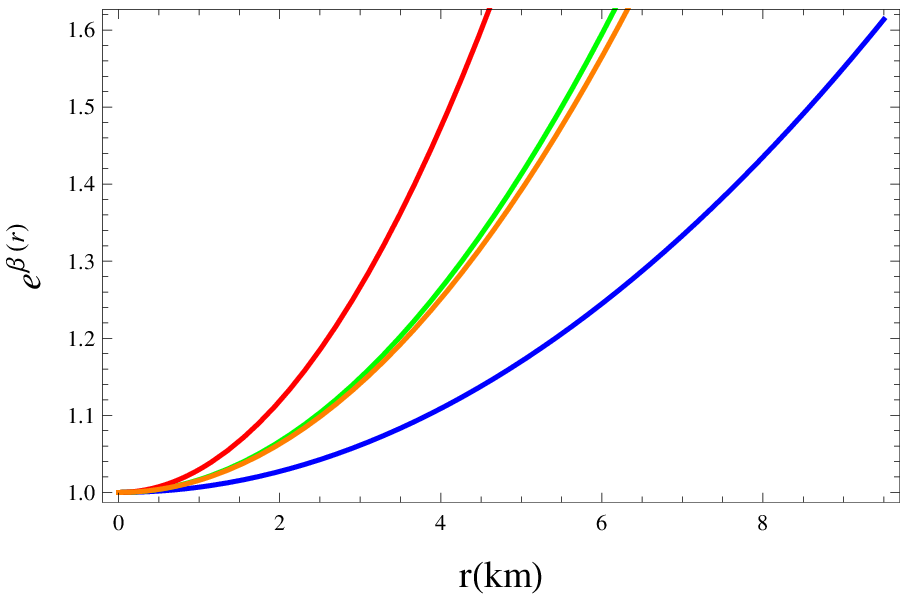,width=.5\linewidth}\caption{Metric potentials
versus radial coordinate.}
\end{figure}

The metric potentials of Finch-Skea symmetry are given as \cite{20}
\begin{eqnarray}\label{11}
e^{\alpha(r)}=(A+\frac{1}{2}Br\sqrt{r^{2}C})^{2}, \quad
e^{\beta(r)}=1+Cr^{2},
\end{eqnarray}
where $A,~B$ and $C$ are constants. These metric potentials gained
the attention of many researchers because they are non-singular and
viable. They have been used to study the compact stellar structures
with various matter distributions \cite{19}-\cite{26}. The graphical
behavior of metric potentials are given in Figure \textbf{1} which
shows that both metric potentials give positively increasing
behavior. We consider $L_{m}=\rho=\frac{m}{V}$,
$P_{r}=\frac{\rho}{3}$, $P_{t}=(\eta+1)P_{r}$, where
$V=\frac{4}{3}\Pi r^{3}$ and $\eta>0$ \cite{51}, hence the field
equations become
\begin{eqnarray}\nonumber
\rho^{eff}&=&\frac{1}{f_{\mathcal{R}}}\bigg[\frac{3m}{4\Pi
r^{3}}-\frac{f-\mathcal{R}f_{\mathcal{R}}}{2}+\frac{3m^{2}}
{4\Pi^{2}r^{6}}(\eta+3)f_{T^{2}}+\frac{1}{1+Cr^{2}}
\\\label{12}
&\times&\bigg\{f''_{\mathcal{R}}-\bigg(\frac{C
r}{1+Cr^{2}}-\frac{2}{r}\bigg)f'_{\mathcal{R}}\bigg\}\bigg],
\\\nonumber
P^{eff}_{r}&=&\frac{1}{f_{\mathcal{R}}}\bigg[\frac{m}{4\Pi
r3}+\frac{1}{2}(f-\mathcal{R}f_{\mathcal{R}})+\frac{m^{2}}{4
\Pi^{2}r^{6}}(1-\eta)f_{T^{2}}-\frac{1}{1+Cr^{2}}
\\\label{13}
&\times&\bigg(\frac{B\sqrt{r^{2}C}}{A+\frac{1}{2}
Br\sqrt{r^{2}C}}+\frac{2}{r}\bigg)f'_{\mathcal{R}}\bigg],
\\\nonumber
P^{eff}_{t}&=&\frac{1}{f_{\mathcal{R}}}\bigg[(1+\eta)\frac{m}{4\Pi
r^{3}}+\frac{1}{2}(f-\mathcal{R}f_{\mathcal{R}})+\frac{m^{2}}{8
\Pi^{2}r^{6}}(2-\eta)f_{T^{2}}
\\\label{14}
&-&\frac{1}{1+Cr^{2}}\bigg\{f''_{\mathcal{R}}+\bigg(\frac{B\sqrt{r^{2}}C}
{A+\frac{1}{2}Br\sqrt{r^{2}C}}-\frac{C
r}{1+Cr^{2}}+\frac{1}{r}\bigg)f'_{\mathcal{R}}\bigg\}\bigg].
\end{eqnarray}
We have used matching conditions to find the values of unknown
constants. These constants explain the configuration and physical
features of compact stars. Goswami et al \cite{52} found that
Schwarzschild spacetime as an exterior region is the best option for
smooth matching. Since EMSG is equivalent to GR in vacuum, hence the
Schwarzschild metric is also a solution for the EMSG field
equations. This is given as
\begin{equation}\label{15}
ds^{2}_{+}=-Ndt^{2}+\frac{1}{N}dr^{2}+r^{2}d
\theta^{2}+r^{2}\sin^{2}\theta d\phi^{2},
\end{equation}
where $N=\frac{r-2M}{r}$ and $M$ defines mass of the star. Using
spacetime continuity at the surface boundary $(r=\mathrm{R})$, we
have
\begin{eqnarray}\label{16}
A=\frac{2\mathrm{R}-5M}{2\sqrt{\mathrm{R}}\sqrt{\mathrm{R}-2M}},
\quad C=\frac{2M}{\mathrm{R}^{2}(\mathrm{R}-2M)}, \quad
B=\frac{\sqrt{\frac{M}{\mathrm{R}-2M}}\sqrt{\mathrm{R}-2M}}{\sqrt{2}\mathrm{R}^{\frac{3}{2}}}.
\end{eqnarray}

\section{Energy-Momentum Squared Gravity Models}

The field equations (\ref{12})-(\ref{14}) appear to be more complex
due to the existence of multivariate functions and their
derivatives. To solve these equations, we consider some specific
models of this theory. For this purpose, we consider the minimal
coupling model of this theory as
\begin{equation}\label{17}
f(\mathcal{R},\mathrm{T}^{2})=f_{1}(\mathcal{R})+f_{2}(T^{2}).
\end{equation}
Different viable models of EMSG can be discussed by considering
various forms of $f_{1}(\mathcal{R})$ with $f_{2}(T^{2})=\gamma
T^{2}$, where $\gamma$ is constant. In the following, we study
various EMSG models corresponding to $f_{1}(\mathcal{R})$.

\subsection*{Model 1}

Here, we use Starobinsky model \cite{54} as
\begin{equation}\label{18}
f(\mathcal{R},\mathrm{T}^{2})=\mathcal{R}+\sigma
\mathcal{R}^{2}+\gamma T^{2},
\end{equation}
where $\sigma$ is greater than or equal to zero. If $\gamma=0$, then
our results reduce to $f(\mathcal{R})$ theory and GR is recovered
for $\sigma=\gamma=0$. Using this model, we have
\begin{eqnarray}\nonumber
\rho^{eff}&=&\frac{1}{1+2\sigma\mathcal{R}}\bigg[\frac{3m}{4\Pi
r^{3}}+\frac{1}{2}\sigma\mathcal{R}^{2}+\frac{\gamma m^{2}}{16
\Pi^{2}r^{6}}\big(30+10\eta-\eta^{2}\big)+\frac{2 \sigma}{1+Cr^{2}}
\\\label{19}
&\times&\bigg\{\mathcal{R}''-\mathcal{R}'\bigg(\frac{C
r}{1+Cr^{2}}-\frac{2}{r}\bigg)\bigg\}\bigg],
\\\nonumber
P^{eff}_{r}&=&\frac{1}{1+2\sigma \mathcal{R}}\bigg[\frac{m}{4\Pi
r^{3}}-\frac{1}{2}\sigma\mathcal{R}^{2}+\frac{\gamma m^{2}}{16
\Pi^{2}r^{6}}\big(10-2\eta+\eta^{2}\big)-\frac{2\sigma
\mathcal{R}'}{1+Cr^{2}}
\\\label{20}
&\times&\bigg(\frac{B\sqrt{r^{2}C}}{A+\frac{1}{2}Br\sqrt{r^{2}C}}+\frac{2}{r}\bigg)\bigg],
\\\nonumber
P^{eff}_{t}&=&\frac{1}{1+2\sigma
\mathcal{R}}\bigg[\big(1+\eta\big)\frac{m}{4\Pi r^{3}}-\frac{1}{2}
\sigma\mathcal{R}^{2}+\frac{\gamma
m^{2}}{16\Pi^{2}r^{6}}\big(10+\eta^{2}\big)
\\\label{21}
&-&\frac{2
\sigma}{1+Cr^{2}}\bigg\{\mathcal{R}''+\mathcal{R}'\bigg(\frac{B\sqrt{r^{2}C}}
{A+\frac{1}{2}Br\sqrt{r^{2}C}}-\frac{C
r}{1+Cr^{2}}+\frac{1}{r}\bigg)\bigg\}\bigg].
\end{eqnarray}

\subsection*{Model 2}

We apply an extension of
$\mathcal{R}-\delta\psi\tanh(\frac{\mathcal{R}}{\delta})$ with
$\delta>0$ and $\psi\geq0$ \cite{55} in the model (\ref{17}) as
\begin{equation}\label{22}
f(\mathcal{R},T^{2})=\mathcal{R}-\psi\delta\tanh(\frac{\mathcal{R}}{\delta})+\gamma
T^{2}.
\end{equation}
The corresponding equations of motion yield
\begin{eqnarray}\nonumber
\rho^{eff}&=&\frac{1}{1-\psi\sec h^{2}\varphi}\bigg[\frac{3m}{4\Pi
r^{3}}+\frac{\psi\delta}{2}\tanh^{2}\varphi-\frac{\psi
\mathcal{R}}{2}\sec h^{2}\varphi
\\\nonumber
&+&\frac{\gamma m^{2}}{16\Pi^{2}r^{6}}\big(30+10\eta-\eta^{2}\big)
+\frac{\psi\sec
h^{4}\varphi}{\delta^{2}(1+Cr^{2})}\bigg\{2\mathcal{R}'^{2}-4\tanh^{2}
\varphi\mathcal{R}'^{2}
\\\label{23}
&+&\delta\sinh2\varphi\bigg(\mathcal{R}''-\mathcal{R}'\bigg(\frac{C
r}{1+Cr^{2}}-\frac{2}{r}\bigg)\bigg)\bigg\}\bigg],
\\\nonumber
P^{eff}_{r}&=&\frac{1}{1-\psi\sec h^{2}\varphi}\bigg[\frac{m}{4\Pi
r^{3}}-\frac{\psi\delta}{2}\tanh\varphi+\frac{\gamma m^{2}}{16
\Pi^{2}r^{6}}\big(10-2\eta+\eta^{2}\big)
\\\nonumber
&+&\psi\sec h^{2}\varphi
\bigg\{\frac{\mathcal{R}}{2}-\frac{2}{\delta(1+Cr^{2})}\tanh\varphi
\mathcal{R}'\bigg(\frac{B\sqrt{r^{2}C}}{A+\frac{1}{2}Br\sqrt{r^{2}C}}
\\\label{24}
&+&\frac{2}{r}\bigg)\bigg\}\bigg],
\\\nonumber
P^{eff}_{t}&=&\frac{1}{1-\psi\sec h^{2}\varphi}\bigg[\big(1+\eta
\big)\frac{m}{4\Pi r^{3}}-\frac{\psi\delta}{2}\tanh
\varphi+\frac{\psi\mathcal{R}}{2}\sec h^{2}\varphi
\\\nonumber
&+&\frac{\gamma m^{2}}{16\Pi^{2}r^{6}}\big(10+\eta^{2}\big)
-\frac{\psi\sec h^{4}
\varphi}{\delta^{2}(1+Cr^{2})}\bigg\{2\mathcal{R}'^{2}-4\tanh^{2}
\varphi \mathcal{R}'^{2}
\\\label{25}
&+&\delta\sinh2\varphi
\bigg(\mathcal{R}''+\bigg(\frac{B\sqrt{r^{2}C}}{A+\frac{1}{2}
Br\sqrt{r^{2}C}}-\frac{C
r}{1+Cr^{2}}+\frac{1}{r}\bigg)\mathcal{R}'\bigg)\bigg\}\bigg],
\end{eqnarray}
where $\varphi=\frac{\mathcal{R}}{\delta}$.

\subsection*{Model 3}

Here, we take an extension of
$[\varepsilon\delta(1+\varphi^{2})^{-S}-1]$ \cite{56} in the model
(\ref{17}) as
\begin{equation}\label{26}
f(\mathcal{R},\mathrm{T}^{2})=\mathcal{R}+\varepsilon\delta
(1+\varphi^{2})^{-S}-1+\gamma T^{2},
\end{equation}
where $\varepsilon$ and $S$ are positive constants. The resulting
equations of motion become
\begin{eqnarray}\nonumber
\rho^{eff}&=&\frac{1}{1-2\varepsilon S
\varphi(1+\varphi^{2})^{-S-1}}\bigg[\frac{3m}{4\Pi
r^{3}}-\frac{\varepsilon
\delta}{2}\big(1+\varphi^{2}\big)^{-S}-\frac{1}{2}-\varepsilon S
\mathcal{R}\varphi
\\\nonumber
&\times&\big(1+\varphi^{2}\big)^{-S-1}+\frac{\gamma
m^{2}}{16\Pi^{2}r^{6}}\big(30+10\eta-\eta^{2}\big)
-\frac{2\varepsilon S(1+ \varphi^{2})^{-S-3}}{\delta^{5}(1+Cr^{2})}
\\\nonumber
&\times&\bigg\{2\mathcal{R}^{2}(-1-S)
(\delta^{2}+\mathcal{R}^{2})\mathcal{R}''+2\mathcal{R}(-1-S)
(3\delta^{2}-2\mathcal{R}^{2}S-\mathcal{R}^{2})
\\\nonumber
&\times&\mathcal{R}'^{2}+(\delta^{2}+\mathcal{R}^{2})\big\{(
\delta^{2}+\mathcal{R}^{2})\mathcal{R}''-\mathcal{R}'\bigg(\frac{C
r}{1+Cr^{2}} -\frac{2}{r}\bigg)
\\\label{27}
&\times&\big(\delta^{2}-2\mathcal{R}^{2}S-\mathcal{R}^{2}\big)
\big\}\bigg\}\bigg],
\\\nonumber
P^{eff}_{r}&=&\frac{1}{1-2\varepsilon S
\varphi(1+\varphi^{2})^{-S-1}}\bigg[\frac{m}{4\Pi
r^{3}}+\frac{\varepsilon
\delta}{2}\big(1+\varphi^{2}\big)^{-S}-\frac{1}{2}+\varepsilon
S\mathcal{R}\varphi
\\\nonumber
&\times&(1+\varphi^{2})^{-S-1}+\frac{\gamma
m^{2}}{16\Pi^{2}r^{6}}\big(10-2\eta+\eta^{2}\big)+\frac{2\varepsilon
S \mathcal{R}'(1+\varphi^{2})^{-S-2}}{\delta^{3}(1+Cr^{2})}
\\\label{28}
&\times&\bigg
(\frac{B\sqrt{r^{2}C}}{A+\frac{1}{2}Br\sqrt{r^{2}C}}+\frac{2}{r}\bigg)
\big(\delta^{2}-\mathcal{R}^{2}-2S\mathcal{R}^{2}\big)\bigg],
\\\nonumber
P^{eff}_{t}&=&\frac{1}{1-2\varepsilon S\varphi(1+
\varphi^{2})^{-S-1}}\bigg[\big(1+\eta\big)\frac{m}{4\Pi
r^{3}}+\frac{\varepsilon
\delta}{2}\big(1+\varphi^{2}\big)^{-S}-\frac{1}{2}
\\\nonumber
&+&\varepsilon S \mathcal{R}
\varphi(1+\varphi^{2})^{-S-1}+\frac{\gamma m^{2}}{16
\Pi^{2}r^{6}}\big(10+\eta^{2}\big)+\frac{2\varepsilon
S(1+\varphi^{2})^{-S-3}}{\delta^{5}(1+Cr^{2})}
\\\nonumber
&\times&\bigg\{2\mathcal{R}^{2}(-S-1)(\delta^{2}+\mathcal{R}^{2})\mathcal{R}''
+2\mathcal{R}(-S-1)(3\delta^{2}-2\mathcal{R}^{2}S-\mathcal{R}^{2})\mathcal{R}'^{2}
\\\nonumber
&+&(\delta^{2}+\mathcal{R}^{2})\big\{(\delta^{2}+\mathcal{R}^{2})
\mathcal{R}''+\mathcal{R}'\bigg(\frac{B\sqrt{r^{2}C}}{A+\frac{1}{2}
Br\sqrt{r^{2}C}}-\frac{C r}{1+Cr^{2}}+\frac{1}{r}\bigg)
\\\label{29}
&\times&(\delta^{2}-2\mathcal{R}^{2}S-\mathcal{R}^{2})\big\}\bigg\}\bigg].
\end{eqnarray}

\section{Physical Characteristics}

This section examines physical characteristics of anisotropic
compact stars graphically. We investigate the behavior of different
physical quantities such as EoS parameters, effective matter
variables, anisotropy, mass, compactness, redshift and energy bounds
in the interior of considered stars. The equilibrium state is
examined by the TOV equation and stability is analyzed via sound
speed and adiabatic index. We use green, red, blue and orange colors
for Vela X-1, SAX J 1808.4-3658, Her X-1 and PSR J0348+0432 stars,
respectively, in all graphs. The values of unknown constants for the
considered compact star models are given in Table \textbf{1}.
\begin{table}\caption{The values of unknown constants for the considered
compact star models.}
\begin{center}
\begin{tabular}{|c|c|c|c|c|c|c|c|}
\hline Compact star models & $\mathrm{M}_{\odot}$ & $\mathrm{R}(km)$
& $\frac{\mathrm{M}}{\mathrm{R}}$
& $\mathrm{A}$ & $\mathrm{B}$  & $\mathrm{C}$ \\
\hline Her X-1 & 0.85 & 8.1 & 0.15478 & 0.7388 & 0.0342867 & 0.00680034\\
\hline  Vela X-1 & 1.77 & 9.05 & 0.29 & 0.43141 & 0.0418945 & 0.0165193\\
\hline  SAX J 1808.4-3658 & 1.435 & 7.07 & 0.31 &  0.400112 & 0.0546312 & 0.0296038\\
\hline  PSR J0348+0432 & 2.1 & 10.06 & 0.29 & 0.374653 & 0.0389364 & 0.0156989\\
\hline
\end{tabular}
\end{center}
\end{table}

\subsection{Effective Fluid Matters and Anisotropy}

The fluid parameters should be maximum at the center of a star and
decrease towards the surface boundary because of the dense profile
of compact stars. Figure \textbf{2} demonstrates that the effective
matter variables behave positively and disappear at the surface
boundary of the stars. Furthermore, their derivatives are negative
and exhibit maximum behavior near the center of the star as given in
Figure \textbf{3}. The graphical behavior indicates that pressure
components and energy density have greater values than GR \cite{57}.
We analyze anisotropic factor to examine the nature of anisotropic
pressure. Anisotropy determines whether pressure is directed inward
or outward. If it is positive then pressure is directed outward,
otherwise pressure is directed inward. According to Figure
\textbf{4}, the profile of anisotropic pressure is positive for all
the selected models, yielding the necessary anti-gravitational force
for compact stellar formations \cite{58}. Moreover, the anisotropy
is increased in EMSG as compared to GR \cite{59}.
\begin{figure}
\epsfig{file=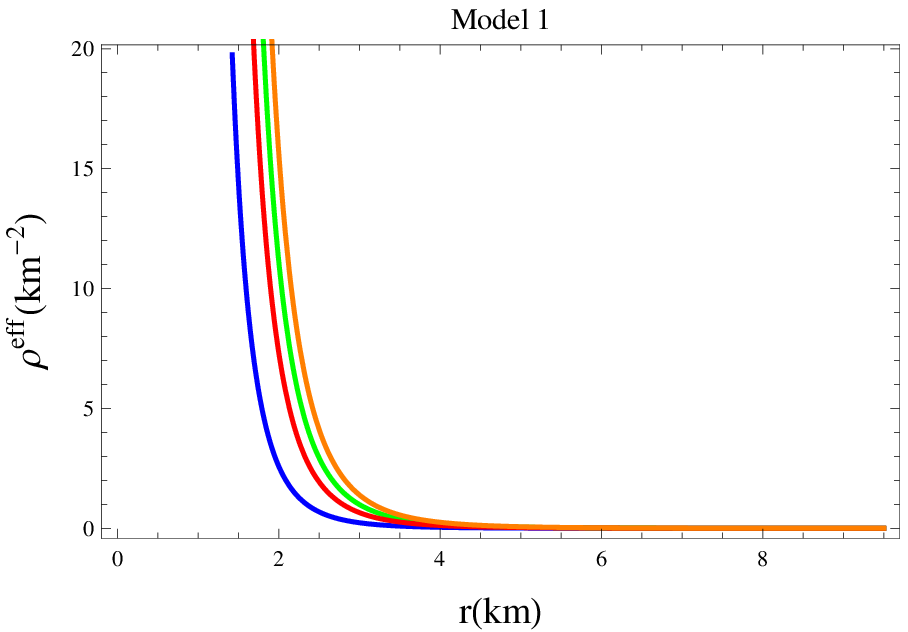,width=.5\linewidth}
\epsfig{file=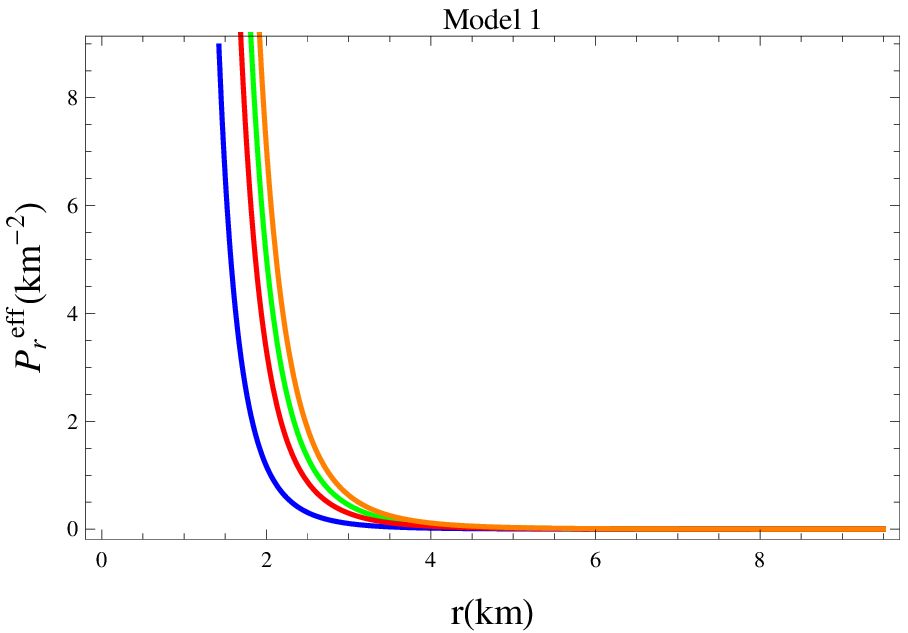,width=.5\linewidth}\center
\epsfig{file=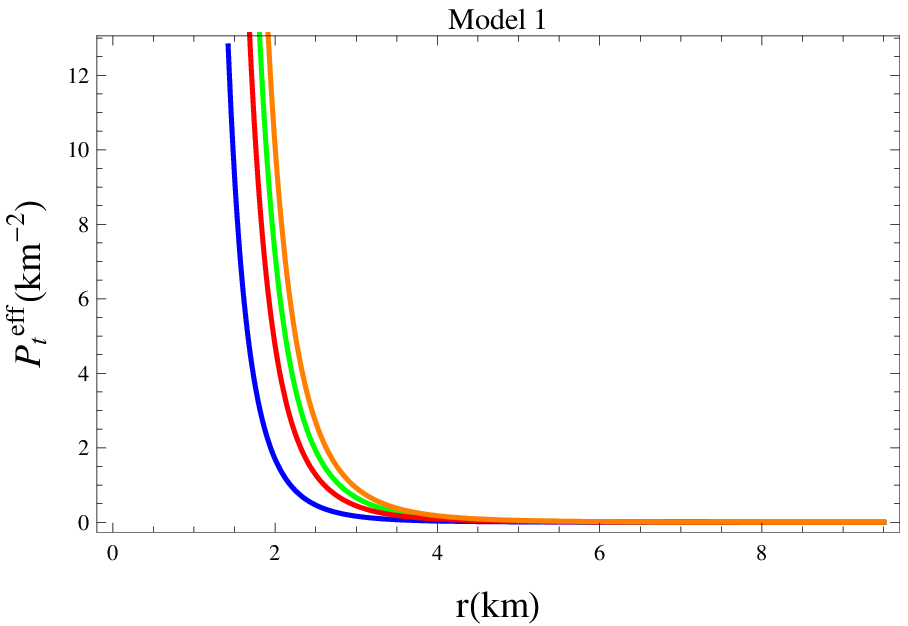,width=.5\linewidth}\caption{Graphs of effective
fluid parameters versus radial coordinate for the model 1.}
\end{figure}
\begin{figure}
\epsfig{file=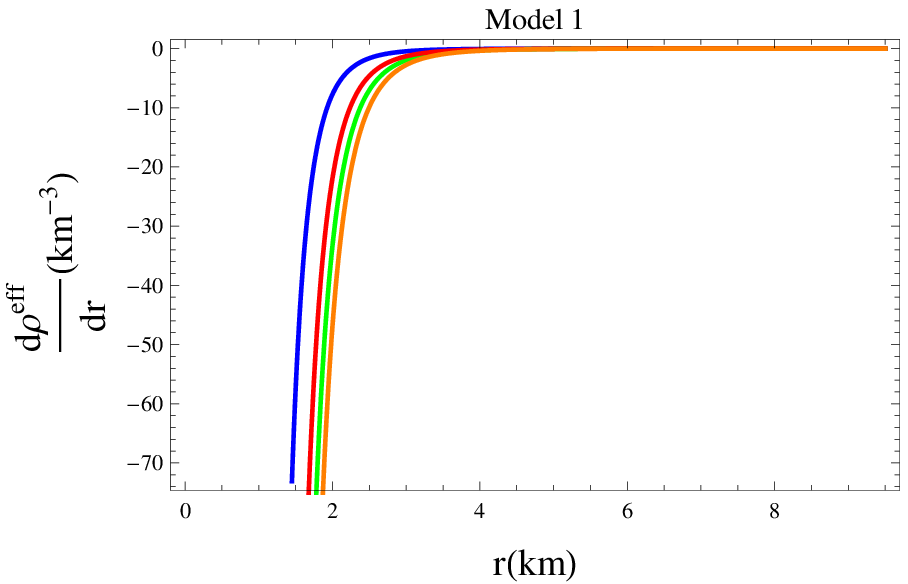,width=.5\linewidth}
\epsfig{file=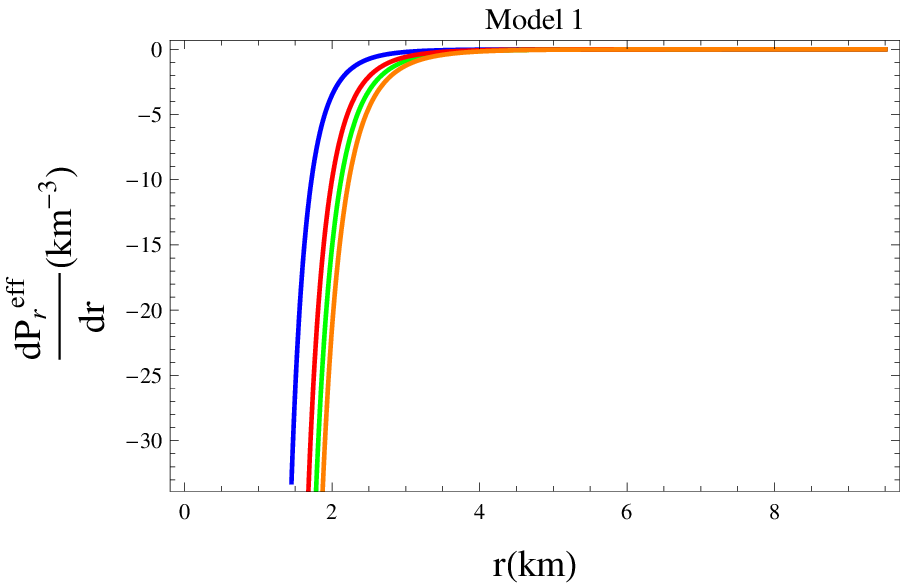,width=.5\linewidth}\center
\epsfig{file=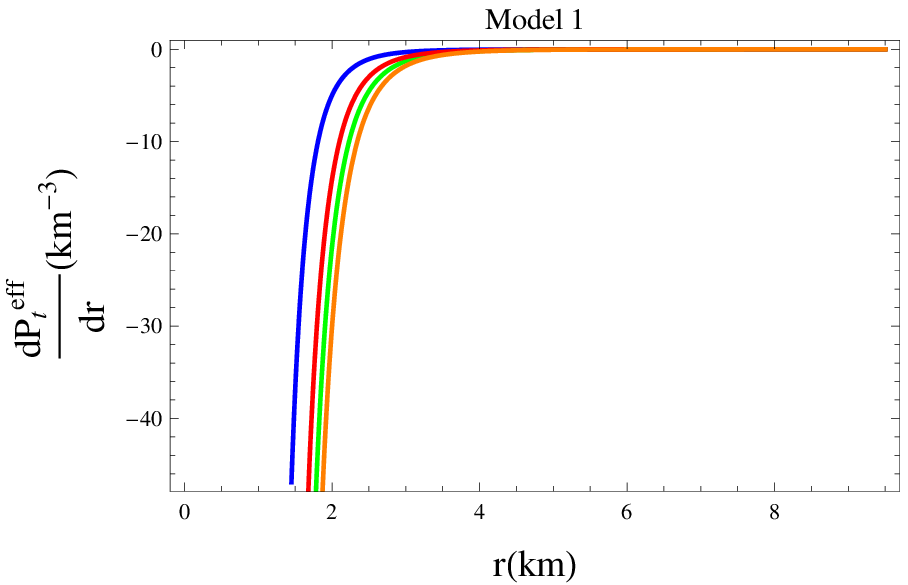,width=.5\linewidth}\caption{Gradients of
effective fluid parameters versus radial coordinate for the models
1.}
\end{figure}

\subsection{Energy Constraints}
\begin{figure}
\center\epsfig{file=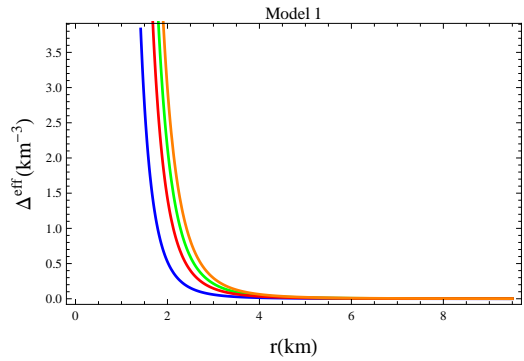,width=.5\linewidth}\caption{Graph of
anisotropy versus radial coordinate.}
\end{figure}

To check the nature of matter (ordinary or exotic), some
mathematical constraints must be imposed on matter, named as energy
constraints. If these constraints are satisfied then ordinary matter
exists in the interior of stars otherwise exotic matter exists.
These energy bounds are classified as
\begin{itemize}
\item null energy condition
\\\\
$\rho^{eff}+P^{eff}_{r}\geq0, \quad \rho^{eff}+P^{eff}_{t}\geq0$,
\item strong energy condition
\\\\
$\rho^{eff}+P^{eff}_{r}\geq0, \quad \rho^{eff}+P^{eff}_{t}\geq0,
\quad \rho^{eff}+P^{eff}_{r}+2P^{eff}_{t}\geq0$,
\item dominant energy condition
\\\\
$\rho^{eff}-P^{eff}_{r}\geq0, \quad \rho^{eff}-P^{eff}_{t}\geq0$,
\item weak energy condition
\\\\
$\rho^{eff}\geq0, \quad \rho^{eff}+P^{eff}_{r}\geq0, \quad
\rho^{eff}+P^{eff}_{t}\geq0$.
\end{itemize}
Figure \textbf{5} demonstrates that all the necessary conditions are
satisfied for the proposed models that indicate the presence of
ordinary matter in the stellar objects.
\begin{figure}
\epsfig{file=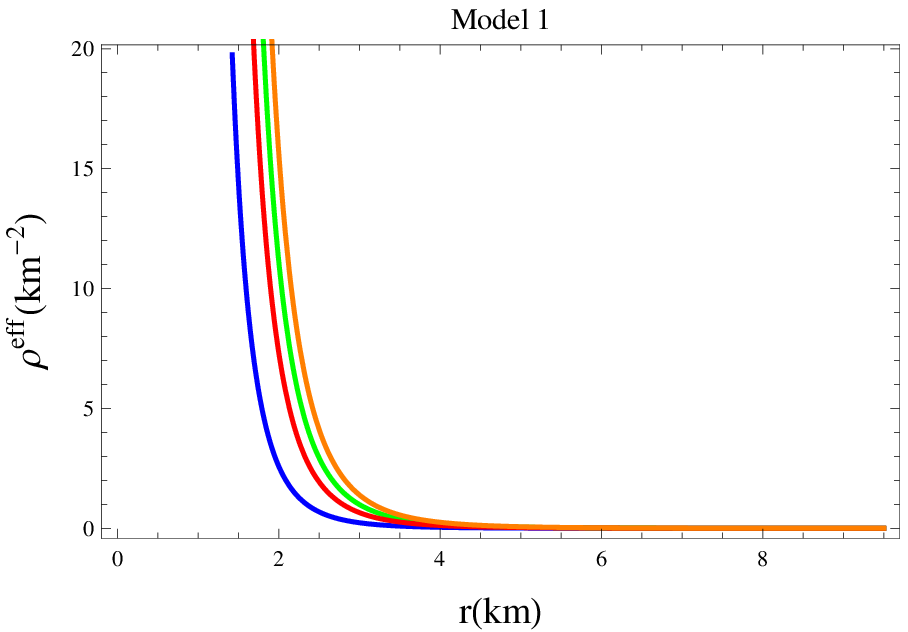,width=.5\linewidth}
\epsfig{file=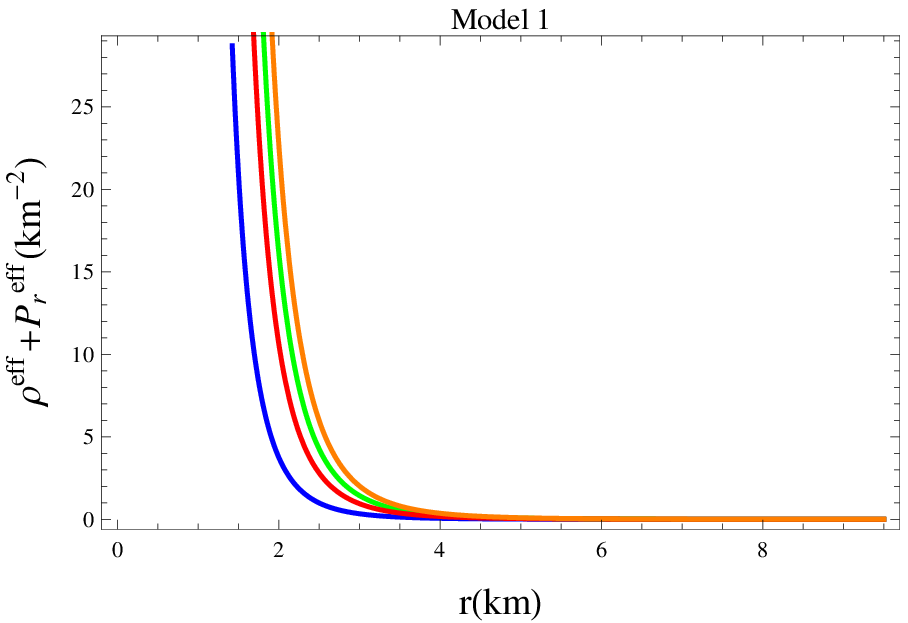,width=.5\linewidth}
\epsfig{file=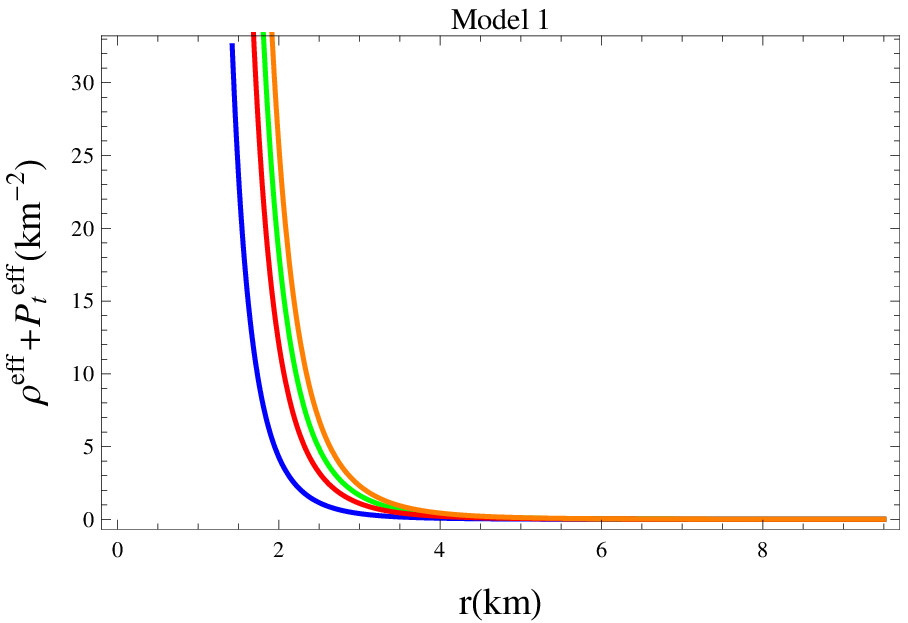,width=.5\linewidth}
\epsfig{file=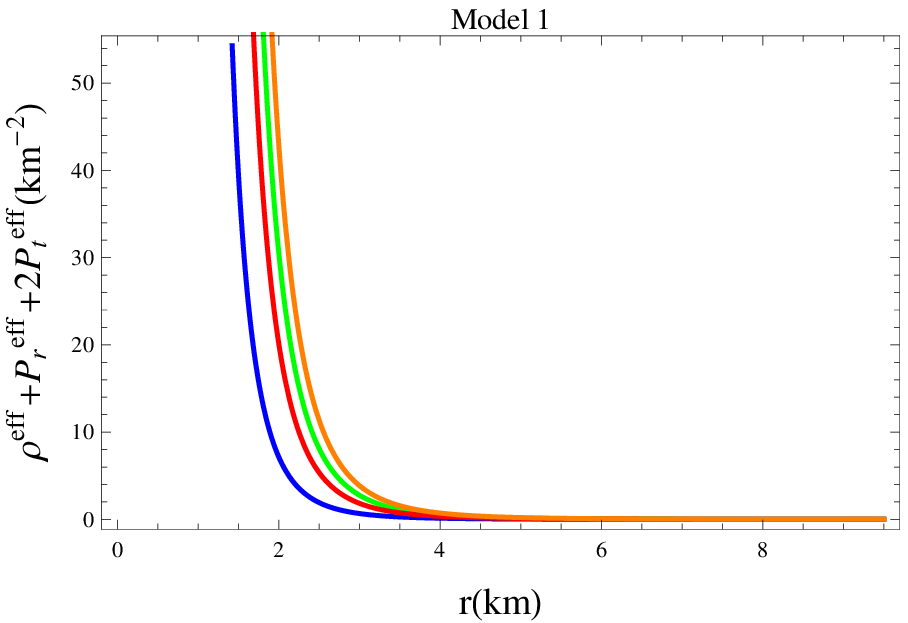,width=.5\linewidth}
\epsfig{file=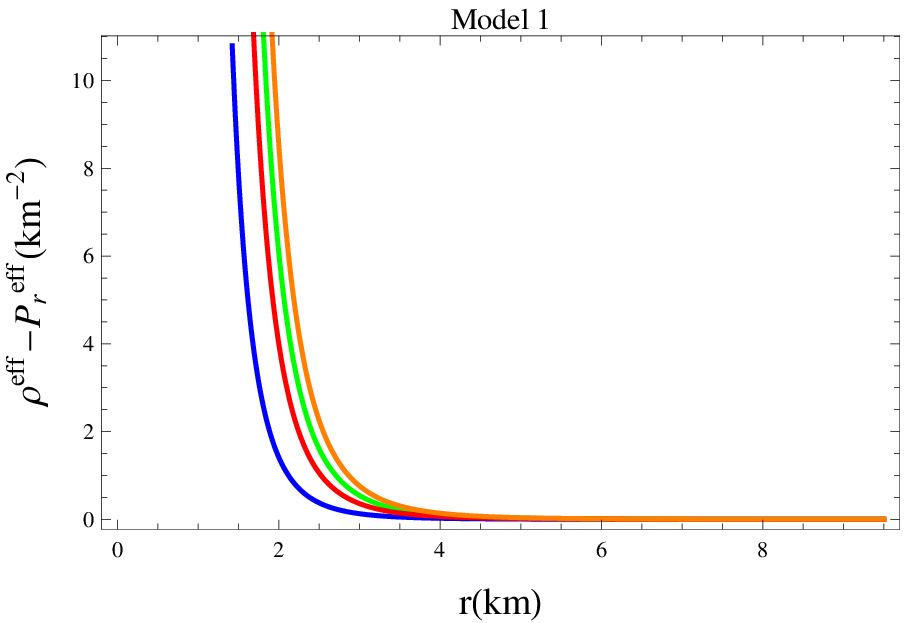,width=.5\linewidth}
\epsfig{file=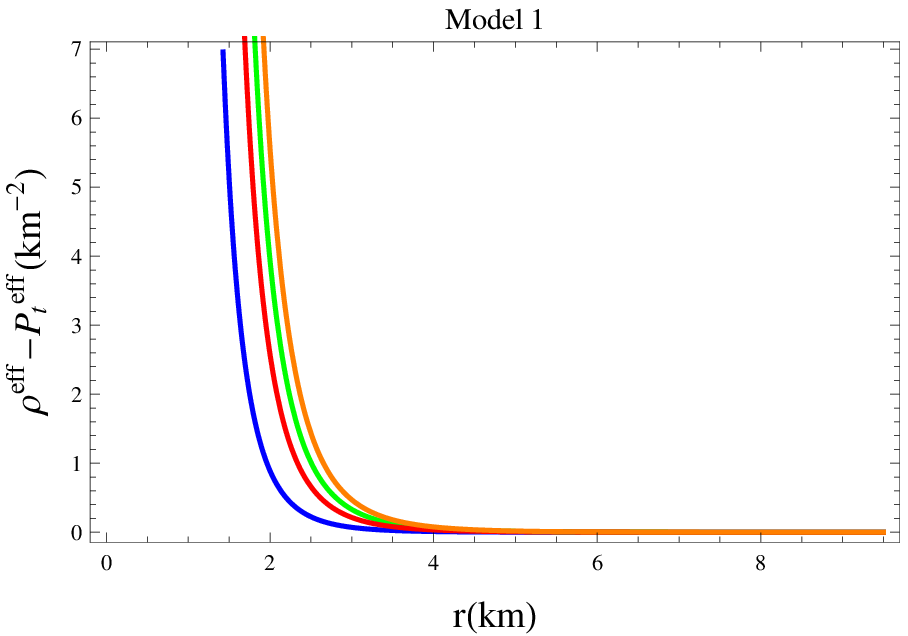,width=.5\linewidth}\caption{Graphs of energy
bounds versus radial coordinate for the model 1.}
\end{figure}

\subsection{Equation of State Parameters}

Here, we examine the EoS parameters which are used to characterize
the relationship among fluid variables. For anisotropic fluid
configuration, the radial and tangential EoS parameters are
expressed as $\omega^{eff}_{r}=\frac{P^{eff}_{r}}{\rho^{eff}}$ and
$\omega^{eff}_{t}=\frac{P^{eff}_{t}}{\rho^{eff}}$, respectively. The
range of EoS parameters should lie in the interval [0,1] for a
physically viable model \cite{44}. Figure \textbf{6} demonstrates
that the EoS parameters for all the models lie between 0 and 1,
indicating the viability of our considered models. It is noted that
EoS parameters have larger values than GR \cite{59}.

\subsection{Mass, Compactness and Redshift}
\begin{figure}
\epsfig{file=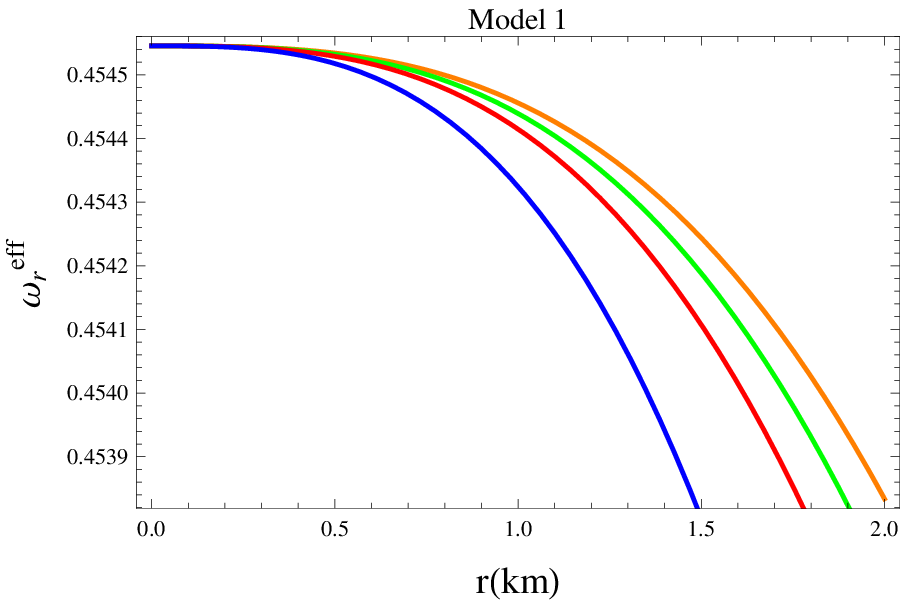,width=.5\linewidth}
\epsfig{file=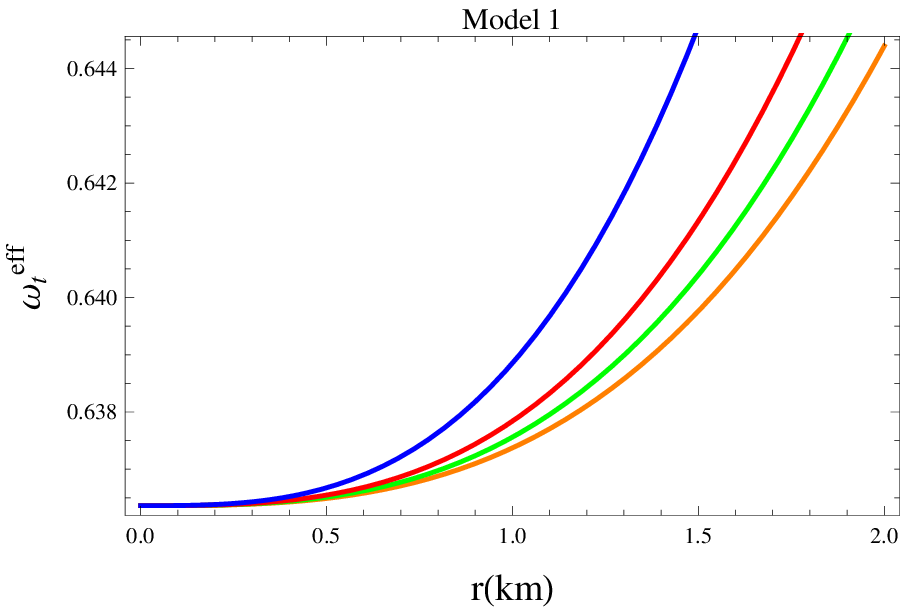,width=.5\linewidth}\caption{Graphs of EoS
parameters versus radial coordinate.}
\end{figure}

The mass of anisotropic compact star is given by
\begin{equation}\label{30}
m=\frac{Mr^{3}e^{\frac{M(r^{2}-R^{2})}{NR^{3}}}}{NR^{3}+2Mr^{2}
e^{\frac{M(r^{2}-R^{2})}{NR^{3}}}}.
\end{equation}
Figure \textbf{7} demonstrates that the mass is positively
increasing with the increase in radius and regular at the center of
the star. To analyze the viability of compact stars, the compactness
function $(u=\frac{m}{r})$ plays a vital role and is expressed as
\begin{figure}
\center\epsfig{file=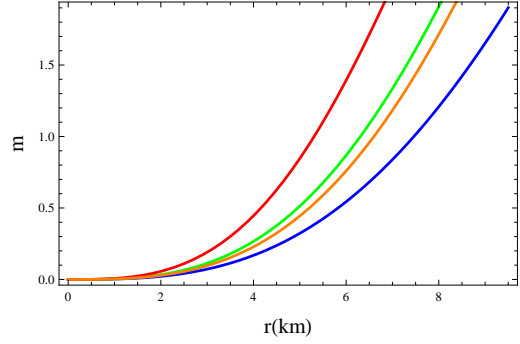,width=.5\linewidth}\caption{Graph of
mass function versus radial coordinate.}
\end{figure}
\begin{equation}\label{31}
u(r)=\frac{Mr^{2}e^{\frac{M(r^{2}-R^{2})}{NR^{3}}}}{NR^{3}+2Mr^{2}
e^{\frac{M(r^{2}-R^{2})}{NR^{3}}}}.
\end{equation}
Buchdahl \cite{60} examined that if this factor possesses the limit
$u(r)<\frac{4}{9}$ then we have viable compact stellar objects. The
gravitational redshift $(Z_{s}=\frac{1}{\sqrt{1-2u}}-1)$ is
considered as the essential term to examine the nature of compact
objects as it calculates the force applied on light due to intense
gravity, described as
\begin{equation}\label{32}
Z_{s}=\frac{1}{\sqrt{1-\frac{2Mr^{2}e^{\frac{M(r^{2}-R^{2})}{NR^{3}}}}
{NR^{3}+2Mr^{2}e^{\frac{M(r^{2}-R^{2})}{NR^{3}}}}}}-1.
\end{equation}
For the viable compact stars, both compactness and redshift must lie
within the specific limits ($u(r)<\frac{4}{9}$ and $Z_{s}\leq5.2$)
\cite{61}. The graphical behavior of compactness and redshift is
given in Figure \textbf{8}, which shows that both are uniformly
increasing and disappear at the center of the star. Also, both
metric potentials satisfy the required limits and hence yield the
viable compact stars in EMSG.
\begin{figure}
\epsfig{file=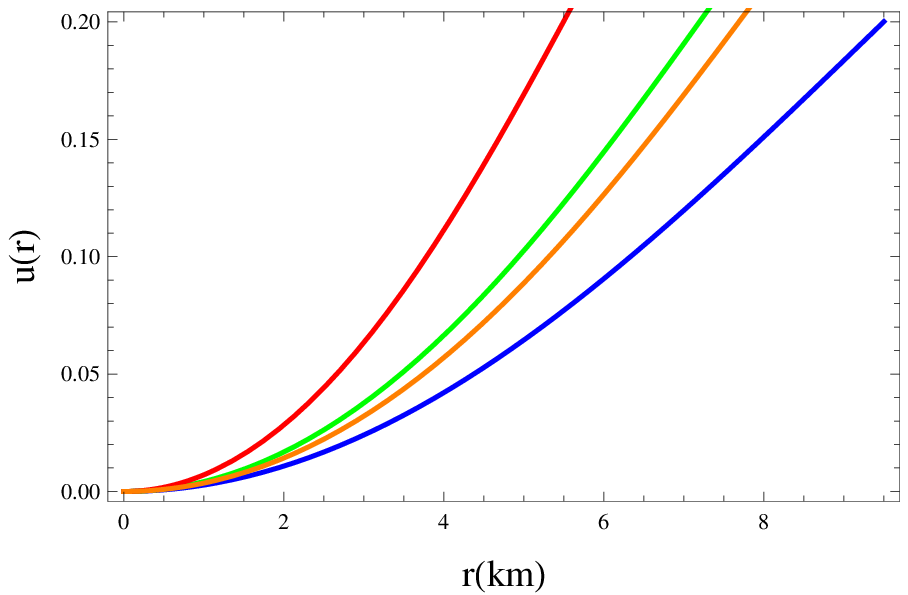,width=.5\linewidth}
\epsfig{file=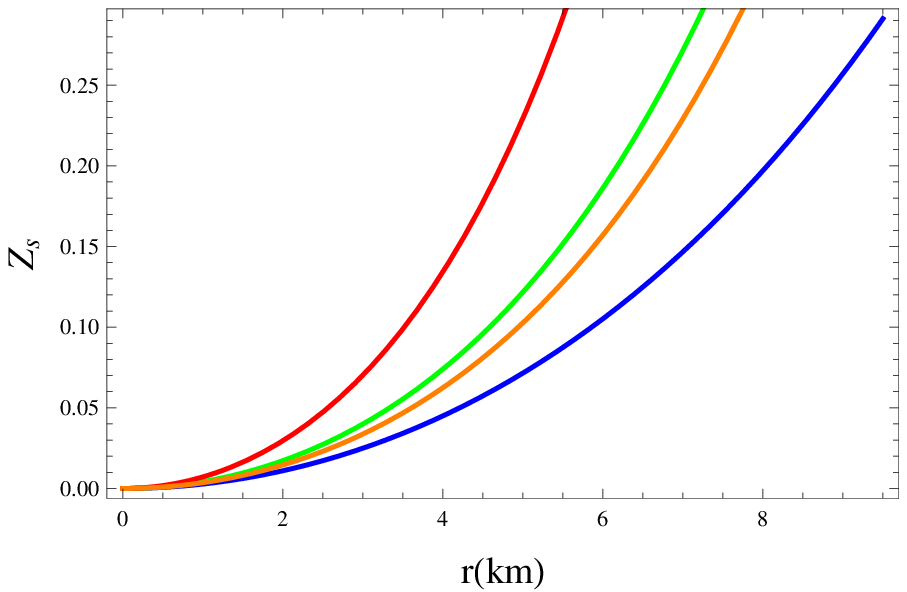,width=.5\linewidth}\caption{Compactness and
redshift functions versus radial coordinate.}
\end{figure}

\section{Equilibrium State and Stability Analysis}

The physically viable models of compact stars depend on the
stability. Such celestial objects that exhibit stable behavior in
the face of external perturbations are more fascinating to observe.
Here, we examine the equilibrium state of our considered star
candidates via TOV equation, and the stability analysis through
sound speed and adiabatic index.

\subsection{TOV Equation}

A mathematical technique that examines the effect of various forces
on compact stellar configuration and reveals their equilibrium
condition is known as TOV equation \cite{62}, defined as
\begin{equation}\label{33}
\frac{(\rho+P^{eff}_{r})M_{G}(r)}{r^{2}}e^{\frac{\alpha-\beta}{2}}
+\frac{2}{r}(P^{eff}_{r}-P^{eff}_{t})+\frac{dP^{eff}_{r}}{dr}=0,
\end{equation}
where $M_{G}(r)$ is the gravitational mass, given by
\begin{equation}\nonumber
M_{G}(r)=4\pi\int(T^{t}_{t}-T^{\theta}_{\theta}-T^{\phi}_{\phi}
-T^{r}_{r})r^{2}e^{\frac{\alpha+\beta}{2}}dr
=\frac{1}{2}e^{\frac{\alpha-\beta}{2}}\alpha'r^{2}.
\end{equation}
Inserting this value in Eq.(\ref{33}), it follows that
\begin{equation}\nonumber
\frac{2}{r}(P^{eff}_{r}-P^{eff}_{t})+(\rho+P^{eff}_{r})\frac{\alpha'}
{2}+\frac{dP^{eff}_{r}}{dr}=0.
\end{equation}
This shows the influence of anisotropic
($F_{a}=\frac{2}{r}(P^{eff}_{r}-P^{eff}_{t})$), gravitational
($F_{g}=(\rho+P^{eff}_{r})\frac{\alpha'}{2}$) and hydrostatic
($F_{h}=\frac{dP^{eff}_{r}}{dr}$) forces on the system. The
graphical behavior of the TOV equation corresponding to Her X-1,
Vela X-1, SAX J 1808.4-3658 and PSR J0348+0432 stars are given in
Figure \textbf{9}. In all TOV graphs, black, pink and yellow lines
represent the gravitational, anisotropic and hydrostatic forces,
respectively. The null effect of these forces gives the equilibrium
state of the stellar objects. Figure \textbf{9} shows that
gravitational, anisotropic and hydrostatic forces cancel each others
effect and exhibit the equilibrium state of the proposed stars.

\subsection{Sound Speed}
\begin{figure}
\epsfig{file=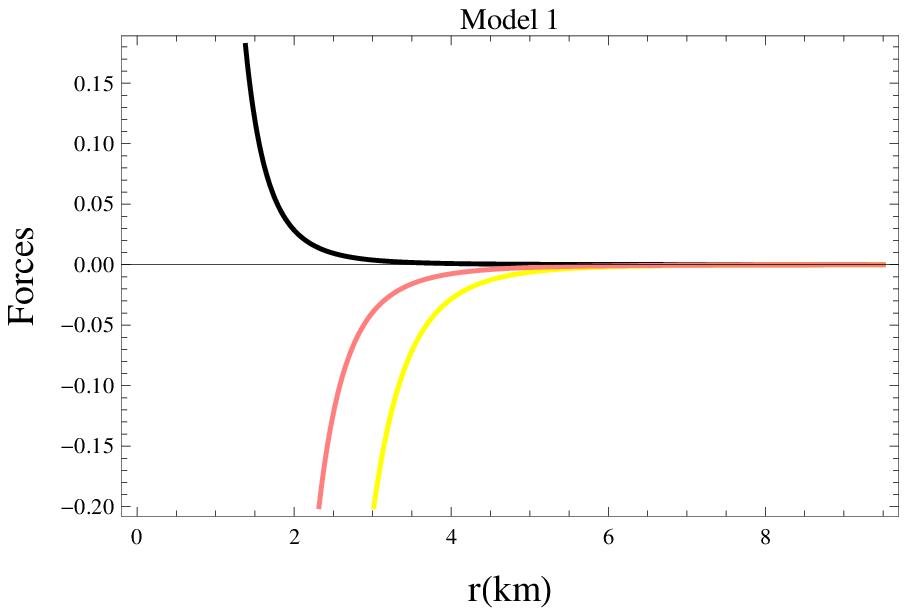,width=.5\linewidth}\epsfig{file=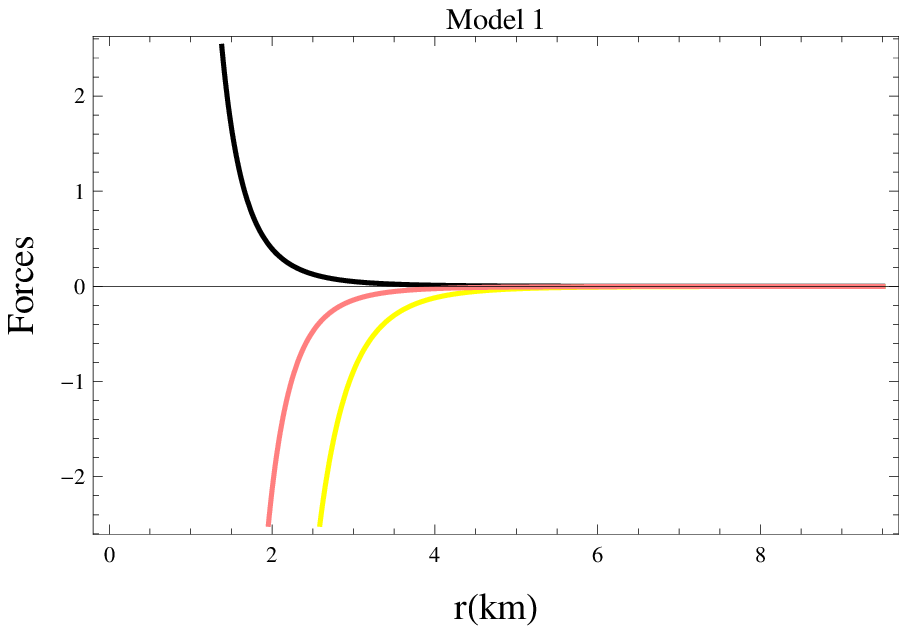,width=.5\linewidth}
\epsfig{file=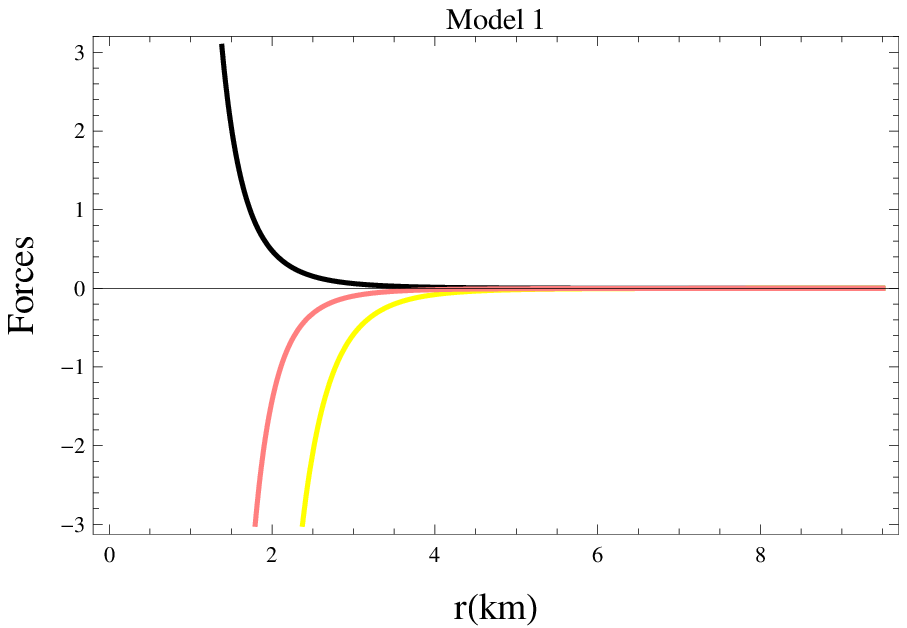,width=.5\linewidth}
\epsfig{file=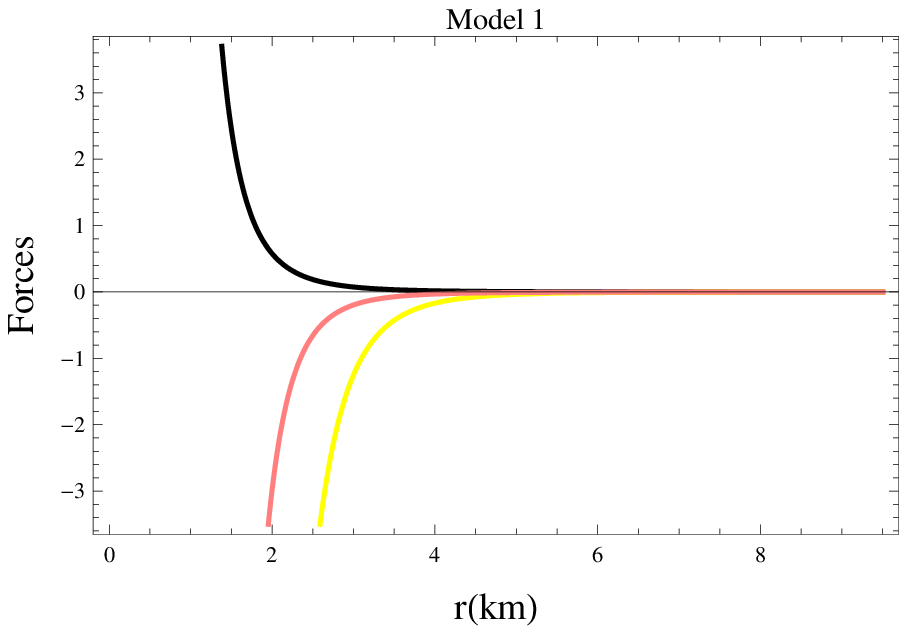,width=.5\linewidth} \caption{Graphs of TOV
equation versus radial coordinate for Her X-1 star (1st), Vela X-1
star (2nd), SAX J 1808.4-3658 star (3rd) and PSR J0348+0432 (4th).}
\end{figure}
\begin{figure}
\epsfig{file=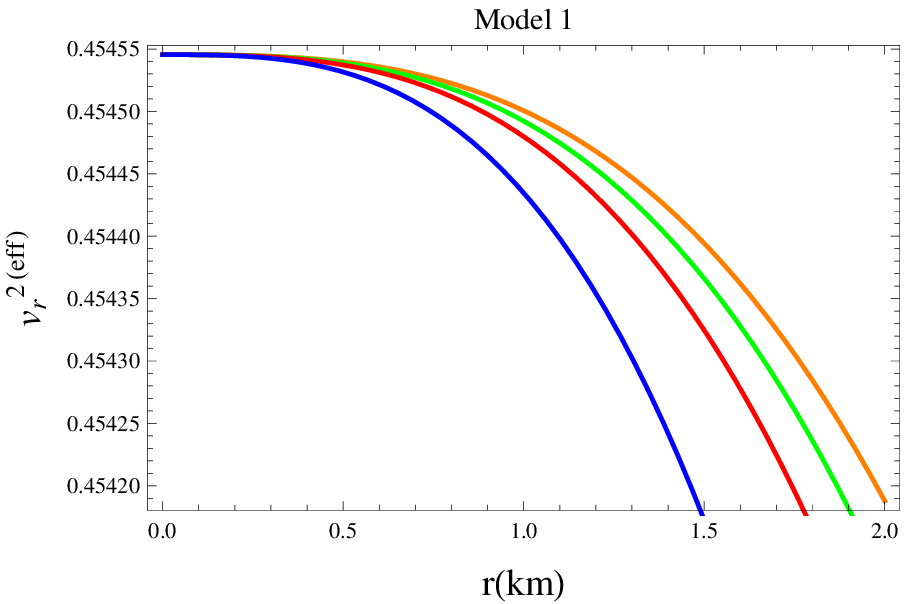,width=.5\linewidth}
\epsfig{file=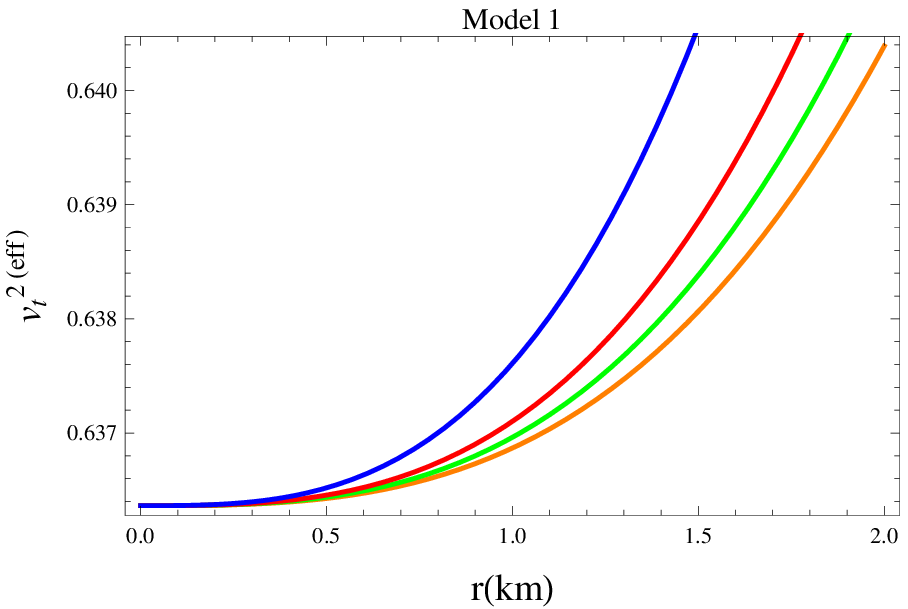,width=.5\linewidth}\center
\epsfig{file=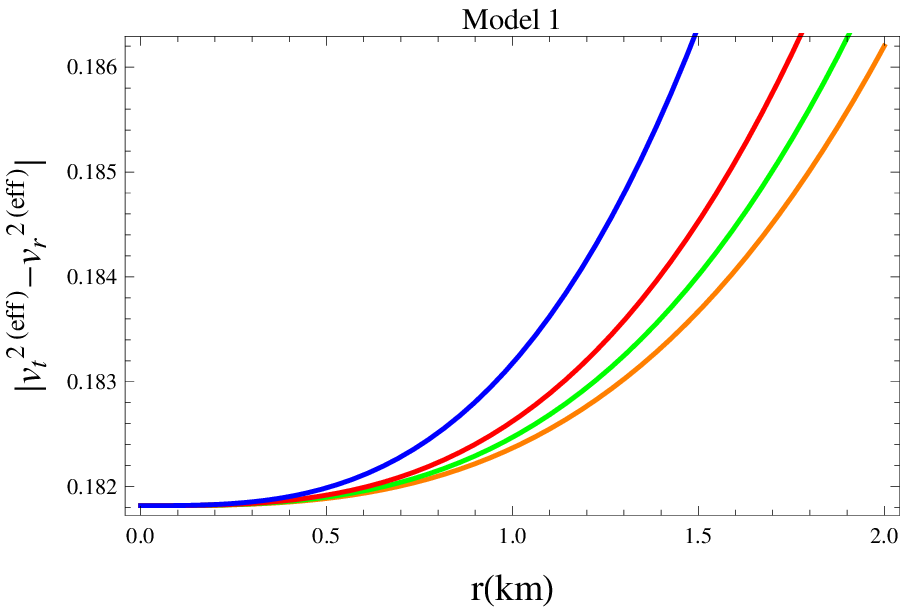,width=.5\linewidth}\caption{Speed of sound
versus radial coordinate for the model 1.}
\end{figure}

Here, we analyze stability of the star candidates through the speed
of sound. This method depends upon the causality criteria and
Herrera cracking technique. The causality criteria states that the
speed of sound components
$(v^{2(eff)}_{sr}=\frac{dP^{eff}_{r}}{d\rho^{eff}},v_{st}^{2(eff)}
=\frac{dP^{eff}_{t}}{d\rho^{eff}})$ should be confined in the range
[0,1] for stable structures. According to Herrera cracking
technique, the difference in sound speed components
$(v^{2(eff)}_{sr}=\frac{dP^{eff}_{r}}{d\rho^{eff}},v_{st}^{2(eff)}
=\frac{dP^{eff}_{t}}{d\rho^{eff}})$ should be $0\leq\mid
v_{st}^{2(eff)}-v^{2(eff)}_{sr}\mid\leq1$. Figure \textbf{10}
demonstrates that the compact stars meet all the necessary criteria
to support the stability of the selected star models of EMSG.

\subsection{Adiabatic Index}

The adiabatic index is an alternative technique to explore the
stability of a compact star. The radial and transverse components of
adiabatic index are expressed as
\begin{eqnarray}\nonumber
\Omega^{eff}_{r}=\frac{\rho^{eff}+P^{eff}_{r}}{P^{eff}_{r}}
\frac{dP^{eff}_{r}}{d\rho^{eff}},
\quad
\Omega^{eff}_{t}=\frac{\rho^{eff}+P^{eff}_{t}}{P^{eff}_{t}}
\frac{dP^{eff}_{t}}{d\rho^{eff}}.
\end{eqnarray}
According to Heintzmann and Hillebrandt \cite{63}, a system is
stable if $\Omega>4/3$, otherwise it is unstable. Figure \textbf{11}
shows that all the proposed compact stars have adiabatic index
components greater than $4/3$, indicating that our system is stable
even when higher-order matter source terms are present.
\begin{figure}
\epsfig{file=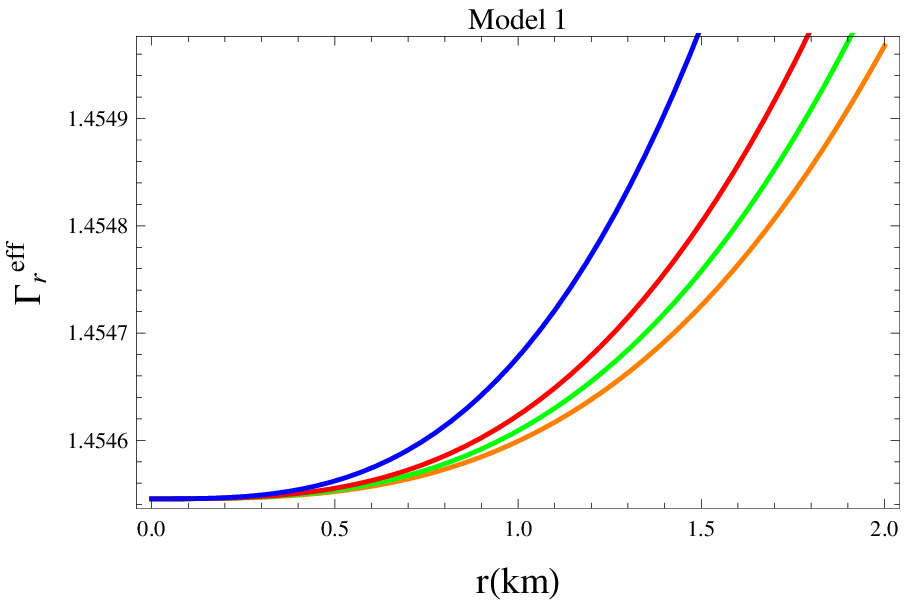,width=.5\linewidth}
\epsfig{file=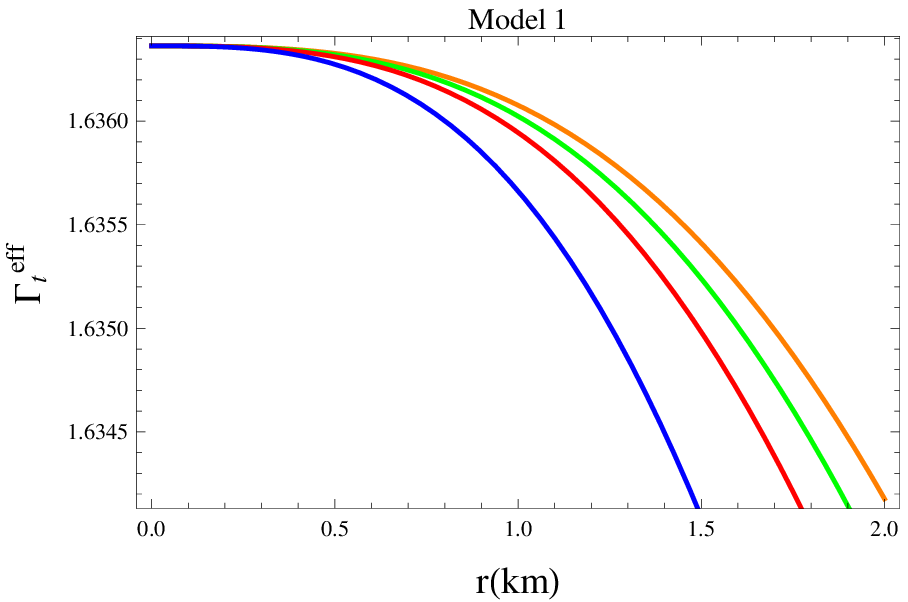,width=.5\linewidth}\caption{Graphs of adiabatic
index versus radial coordinate.}
\end{figure}

\section{Final Remarks}

This paper examines various physically viable and stable compact
star models using Finch-Skea symmetry with anisotropic matter
compositions in $f(\mathcal{R},\mathrm{T}^{2})$ theory. We have used
observational values of the radius and mass of these massive stars
to find the values of unknown constants. We have explored the
graphical behavior of fluid variables, energy conditions and EoS
parameters inside compact stars using different
$f(\mathcal{R},\mathrm{T}^{2})$ models. Finally, we have discussed
the equilibrium state of stellar objects by TOV equation and their
stability is examined through causality condition, Herrera cracking
approach and the adiabatic index.

We have found that both metric potentials are consistent and satisfy
the necessary conditions, i.e., they have smallest value at the core
of stars and thereafter show monotonically increasing behavior
(Figure \textbf{1}). The effective fluid parameters show maximum
value near the center of the stars and uniformly decrease
corresponding to $r$ (Figures \textbf{2}). The derivatives of
effective matter variables are negative which is required for the
considered stars (Figures \textbf{3}). The nature of anisotropy is
positive which indicates the repulsive anisotropic force necessary
for compact star objects (Figure \textbf{4}). All energy constraints
are satisfied indicating that ordinary matter exists in the interior
of stellar objects (Figures \textbf{5}). The range of EoS parameters
lies in the interval [0,1], indicating viability of the considered
models (Figure \textbf{6}). The mass function is positively
increasing and depends upon the radial coordinate (Figure
\textbf{7}). The behavior of compactness and redshift increases and
satisfies the required conditions (Figure \textbf{8}). The
equilibrium as well as stability requirements are satisfied in the
presence of modified  $f(\mathcal{R},\mathrm{T}^{2})$ terms (Figures
\textbf{9-11}).

It is found that values of all the considered physical parameters
increase in EMSG as compared to GR \cite{57,59} and other modified
theories \cite{44,64}. We conclude that the compact stars are
physically viable and stable in $f(\mathcal{R},\mathrm{T}^{2})$
theory.\\\\
\textbf{Data Availability:} This manuscript has no associated data.

\end{document}